\documentclass[
reprint,
superscriptaddress,
nobibnotes,
 amsmath,
 amssymb,
 aps,
prl,
floatfix,
]{revtex4-1}

\usepackage{graphicx}
\usepackage{dcolumn}
\usepackage{bm}
\usepackage{physics}
\usepackage{siunitx}
\usepackage{siunitx}
\usepackage{booktabs}
\usepackage{multirow}

\usepackage{makecell} 
\usepackage{booktabs} 
\usepackage{mathtools} 
\DeclareSIUnit\angstrom{\text{Å}}

\usepackage{braket}

\usepackage{acro} 
    \DeclareAcronym{PL}{
      short = PL,
      long  = photoluminescence 
    }
    \DeclareAcronym{PLE}{
      short = PLE,
      long  = Photoluminescence excitation 
    }
    \DeclareAcronym{FWHM}{
      short = FWHM,
      long  = full width at half maximum 
    }
    \DeclareAcronym{TM}{
      short = TM,
      long  = transverse magnetic 
    }
    \DeclareAcronym{TE}{
      short = TE,
      long  = transverse electric  
    }
    \DeclareAcronym{PIC}{
      short = PIC,
      long  = photonic integrated circuit 
    }
    \DeclareAcronym{PDMS}{
      short = PDMS,
      long  = polydimethylsiloxane 
    }
    \DeclareAcronym{PC}{
      short = PC,   
      long  = polycarbonate 
    }
    \DeclareAcronym{SOI}{
      short = SOI,   
      long  = silicon-on-insulator 
    }
    \DeclareAcronym{SPhPs}{
      short = SPhPs,   
      long  = surface phonon polaritons 
    }
    \DeclareAcronym{REI}{
      short = REI,   
      long  = Rare-earth ion 
    }
\usepackage[colorlinks=true, linkcolor=blue, citecolor=blue, urlcolor=blue]{hyperref}
\usepackage{orcidlink}

\begin{document}
\setcounter{secnumdepth}{1}

\title{Optical Spectroscopy of Waveguide coupled Er$^{3+}$ ensembles in CaWO$_4$ and YVO$_4$}

\author{Fabian Becker\,\orcidlink{0000-0002-4447-4211}}
    \email{fa.becker@tum.de}
    \affiliation{TUM School of Computation, Information and Technology, Technical University of Munich, 80333 Munich, Germany}%
    \affiliation{Walter Schottky Institut, Technical University of Munich, 85748 Garching, Germany}%
    \affiliation{Munich Center for Quantum Science and Technology (MCQST), 80799 Munich, Germany}%
\author{Anna Selzer\,\orcidlink{0009-0001-7652-5044}}
    \affiliation{TUM School of Computation, Information and Technology, Technical University of Munich, 80333 Munich, Germany}%
    \affiliation{Walter Schottky Institut, Technical University of Munich, 85748 Garching, Germany}%
    \affiliation{Munich Center for Quantum Science and Technology (MCQST), 80799 Munich, Germany}%
\author{Lorenz J. J. Sauerzopf\,\orcidlink{0009-0001-7652-5044}}
    \affiliation{TUM School of Computation, Information and Technology, Technical University of Munich, 80333 Munich, Germany}%
    \affiliation{Walter Schottky Institut, Technical University of Munich, 85748 Garching, Germany}%
    \affiliation{Munich Center for Quantum Science and Technology (MCQST), 80799 Munich, Germany}%
\author{Catherine L. Curtin}
    \affiliation{TUM School of Computation, Information and Technology, Technical University of Munich, 80333 Munich, Germany}%
    \affiliation{Walter Schottky Institut, Technical University of Munich, 85748 Garching, Germany}%
    \affiliation{Munich Center for Quantum Science and Technology (MCQST), 80799 Munich, Germany}%
\author{Sudip KC\, \orcidlink{0009-0000-6269-4763}}
    \affiliation{TUM School of Computation, Information and Technology, Technical University of Munich, 80333 Munich, Germany}%
    \affiliation{Walter Schottky Institut, Technical University of Munich, 85748 Garching, Germany}%
    \affiliation{Munich Center for Quantum Science and Technology (MCQST), 80799 Munich, Germany}%
\author{Tim Schneider\, \orcidlink{0009-0000-4096-2237}}
    \affiliation{TUM School of Computation, Information and Technology, Technical University of Munich, 80333 Munich, Germany}%
    \affiliation{Walter Schottky Institut, Technical University of Munich, 85748 Garching, Germany}%
    \affiliation{Munich Center for Quantum Science and Technology (MCQST), 80799 Munich, Germany}%
\author{Kai M\"uller\, \orcidlink{0000-0002-4668-428X}}
    \email{kai.mueller@tum.de}
    \affiliation{TUM School of Computation, Information and Technology, Technical University of Munich, 80333 Munich, Germany}%
    \affiliation{Walter Schottky Institut, Technical University of Munich, 85748 Garching, Germany}%
    \affiliation{Munich Center for Quantum Science and Technology (MCQST), 80799 Munich, Germany}%

\date{\today}

\begin{abstract}
We present an optical study of near-surface Er$^{3+}$ ensembles in waveguide-integrated CaWO$_4$ and YVO$_4$, investigating how nanophotonic coupling modifies rare-earth spectroscopy. In particular, we compare bulk excitation with evanescently coupled TE and TM waveguide modes. In Er$^{3+}$:CaWO$_4$, we observe a pronounced polarization-dependent surface effect. TE-coupled spectra closely reproduce bulk behavior. In contrast, TM coupling induces strong inhomogeneous broadening and an asymmetric low-energy shoulder of the site S1 Y1Z1 transition, with linewidths exceeding those of the bulk by more than a factor of four. Temperature-dependent measurements and surface termination studies indicate that surface charges are the dominant mechanism. Er$^{3+}$:YVO$_4$ remains largely unaffected by mode polarization, and surface termination leads only to minor spectral shifts. These observations suggest that non-charge-neutral rare-earth systems are more susceptible to surface-induced decoherence sources than charge-neutral hosts.
\end{abstract}
\maketitle
\section{Introduction} \label{S_Introduction}
\ac{REI} doped crystals have emerged as a promising solid-state platform for quantum technologies due to their combination of narrow optical transitions, long spin coherence times, and, in the case of Er$^{3+}$, compatibility with telecom wavelengths~\cite{Stevenson.2022, Gritsch.2022, Reiserer.2022}. These properties enable coherent spin–photon interfaces that are essential for quantum communication, quantum memories, and microwave-to-optical transduction~\cite{Xie.2025}. In addition, the atomic nature of \ac{REI}s provides highly stable and reproducible optical frequencies, supporting scalable quantum networks with intrinsic spectral matching between remote nodes~\cite{Uysal.6102024}.\\
Integrating rare-earth–doped systems into on-chip nanophotonic devices allows realizing a large number of functional units within a small footprint, which is essential for scaling quantum architectures beyond individual bulk experiments~\cite{Rinner.2023}. Moreover, compared to macroscopic bulk setups, on-chip integration reduces the cooling power required per device, enabling the operation of many elements in parallel under cryogenic conditions.\\
Several approaches have been explored to integrate rare-earth–doped crystals into nanophotonic devices, demonstrating the versatility of this platform for on-chip quantum technologies. Early efforts included the fabrication of amorphous-silicon photonic resonators directly on Er$^{3+}$:Y$_2$SiO$_5$ substrates~\cite{Miyazono.2017}, providing a first hybrid on-chip interface but with limited optical performance. More recently, transfer printing of pre-fabricated silicon photonic circuits onto \ac{REI} host crystals has emerged as a powerful method to combine high-quality nanophotonic structures with well-controlled rare-earth environments. This technique has been applied to Er$^{3+}$:Y$_2$SiO$_5$~\cite{Dibos.2018, Okajima.2025} and extended to alternative hosts such as Er$^{3+}$:CaWO$_4$~\cite{Ourari.192023, Uysal.6102024} and Er$^{3+}$:MgO~\cite{Horvath.762023}. In parallel, direct nanostructuring within the \ac{REI} host materials Y$_2$SiO$_5$ and YVO$_4$ has been realized, enabling cavity-enhanced emission and coherent control of individual ions~\cite{Craiciu.2019, Bartholomew.2020, Kindem.2020}. Complementary approaches implant Er$^{3+}$ directly into silicon, establishing telecom-band spin–photon interfaces compatible with integrated photonic platforms~\cite{Gritsch.2025, Gritsch.2022}.\\
For this study, we use a transfer printing approach to place \ac{SOI} \ac{PIC}s on top of complex crystals. With these transferred structures, we investigate Er$^{3+}$:CaWO$_4$ as a promising platform for quantum applications and use Er$^{3+}$:YVO$_4$ as a comparison system. Er$^{3+}$:CaWO$_4$ exhibits exceptionally long spin coherence times, reaching $\SI{23}{ms}$ under millikelvin conditions due to its low nuclear-spin density and weak magnetic noise environment~\cite{LeDantec.2021}. This makes it one of the most coherent natural-abundance materials reported to date. In this host, Er$^{3+}$ substitutes for Ca$^{2+}$ in an S$_4$ site in a non–charge-neutral configuration, which requires charge compensation mainly of long range nature~\cite{Becker.2025, Becker.10242025}. Additionally, the S$_4$ site consists of two inversion symmetry-related sites, which can spectroscopically be separated by an applied electric field~\cite{Billaud.2025,Mims.1965}. In contrast, Er$^{3+}$ substitutes isovalently for Y$^{3+}$ in YVO$_4$, occupying a highly symmetric D$_{2d}$ site without the need for charge compensation~\cite{Xie.2021}, resulting in narrow optical transitions and well-understood spin and optical properties. While the bulk characteristics of both systems are well established \cite{Becker.2025, Xie.2021}, their surface behavior remains largely unexplored. Recent microwave studies have probed near-surface Er$^{3+}$ ions in CaWO$_4$, revealing strain-induced line distortions and modified spin properties near metal interfaces~\cite{Billaud.2025}. Yet, a systematic optical investigation of rare-earth ions in proximity to a surface, where dynamic surface effects can broaden or shift the transitions, has not been performed.\\
Here, we present a comparative optical study of waveguide coupled, natural abundant Er$^{3+}$:CaWO$_4$ ($\SI{10}{ppm}$) and Er$^{3+}$:YVO$_4$ ($\SI{200}{ppm}$), focusing on the mechanisms of spectral broadening and environmental perturbation. The fundamental difference in charge neutrality and symmetry between these two hosts provides an ideal framework to separate intrinsic surface effects from host-dependent contributions.

\section{Device Fabrication, Design and Performance} \label{S_Design}
\begin{figure*}[t]
    \centering
    \includegraphics[page=3, width=\linewidth]{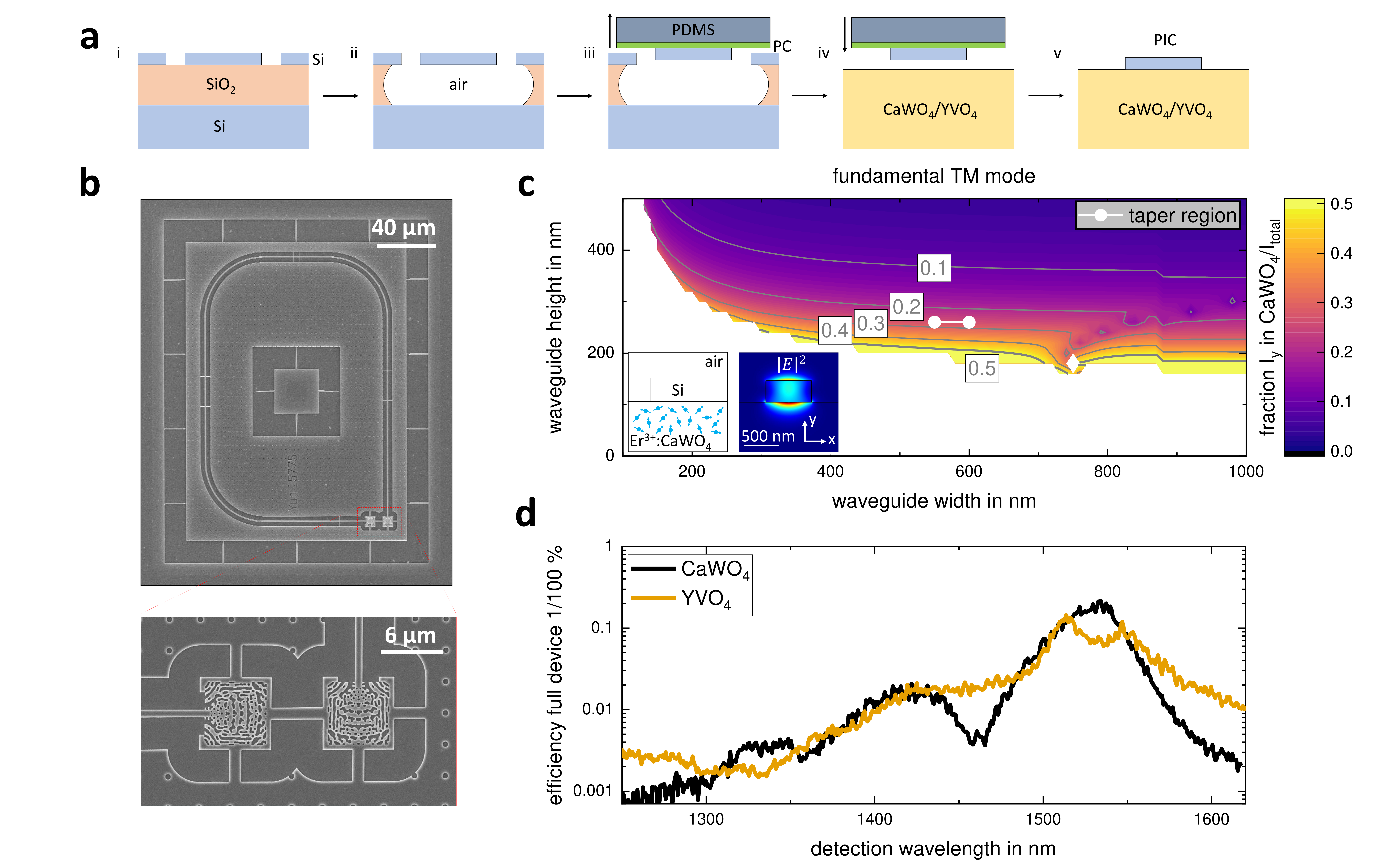}
    \caption{Device transfer printing, design, and performance. \textbf{a} Schematic of transfer printing from patterned Si device layer (i), across wet chemical HF under-etching and critical point drying (ii), towards the actual \ac{PIC} harvest (iii), release (iv) and \ac{PC} removal (v) \textbf{b} SEM image of a looped waveguide hung inside a suspended support membrane with inverse-designed grating couplers optimized for normal free-space to fundamental TM mode coupling. \textbf{c} Fraction of the y-polarized light intensity inside CaWO$_4$/total intensity as a function of the waveguide design. Insets show the total TM mode intensity for the applied design (width $=\SI{650}{nm}$, height $\SI{550}{nm}$). \textbf{d} Example transmission efficiencies of full devices.}
    \label{F_Figure_01}
\end{figure*}

For the transfer printing of \ac{PIC}s, we relied on the method shown by Dibos et al. (2018)~\cite{Dibos.2018}. In contrast, we used a \ac{PC}/\ac{PDMS} stamp, as it turned out that we have more control during the device harvest and release steps compared to a \ac{PDMS} stamp only. The fabrication process is highlighted in Figure~\ref{F_Figure_01}a. It schematically illustrates (i) the patterned \ac{PIC} inside a \ac{SOI} chip after e-beam lithography, cryogenic reactive ion etching, and removal of the photoresist. The designs are connected by bridges to the surrounding silicon layer. In the next step (ii), we put the sample inside a $\SI{10}{\%}$ concentrated HF bath, dip it in deionized water, and remove liquids inside a critical point dryer. As a result, we receive chips with hundreds of freestanding membranes containing \ac{PIC}s. For the harvest (iii), we bring our prepared \ac{PC}/\ac{PDMS} stamp in contact with the desired devices at $\SI{90}{\degree C}$ and pick these membranes up. For the release (iv), we first bring the \ac{PIC} membranes and crystal into contact. Subsequently, we melt the \ac{PC} at a sample temperature of $\SI{190}{\degree C}$ such that \ac{PC} and membrane remain on top of the crystal. As a final step (v), we remove the \ac{PC} in a chloroform bath and apply an ultrasonic cleaning bath in isopropanol.\\
An example of a freestanding \ac{PIC} inside a support membrane is visible in the SEM image in Figure~\ref{F_Figure_01}b. The \ac{PIC} consists of a simple loop with two inverse design grating couplers on both ends. It includes a support membrane connected to tipped bridges to the outer silicon layer and an inside square pillar. The waveguide itself is attached to the support lamella with a few $\SI{100}{nm}$ thick bridges. We chose two grating couplers placed close together to, on the one hand, see if laser light is transmitted through the \ac{PIC}, and on the other hand, fit both grating couplers within the field of view of our cryogenic setup. The grating couplers are optimized to support the fundamental \ac{TM} mode. For CaWO$_4$ the waveguide at the grating coupler is $\SI{260}{nm}$ high and $\SI{550}{nm}$ wide and adjusts via an adiabatic taper to a width of $\SI{600}{nm}$ to reduce bending losses in the $\SI{40}{\micro\meter}$ radius bends. For YVO$_4$, the waveguide tapers from $\SI{750}{nm}$ to $\SI{800}{nm}$ given its higher refractive index.
We chose the \ac{TM} mode to align with the electric field normal to the crystal surface, and thus, coupling merely with the same dipole orientation across the waveguide design. Additionally, the \ac{TM} mode penetrates the crystal deeper compared to the \ac{TE} mode, which allows deeper localized Er$^{3+}$ ions to be involved in the coupling. From a practical point of view, these deeper ions should experience less distortion by the surface. For CaWO$_4$ the \ac{PIC}s are placed in the (100) plane, while for YVO$_4$ the \ac{PIC}s are placed in the (001) plane. Figure~\ref{F_Figure_01}c shows the fraction of y-polarized intensity inside CaWO$_4$ divided by the total intensity as a function of the waveguide height and width, and its insets highlight the involved mode profile and design. As a trade-off between higher loss at the cut-off and low evanescent coupling, we chose a height of $\SI{260}{nm}$ and a width between $\SI{550}{nm}$ and $\SI{600}{nm}$. With these waveguide parameters, we expect about $\SI{27}{\%}$ of the entire intensity to be y-polarized and located in the CaWO$_4$ crystal.\\
The grating couplers are topologically inverse designed with a footprint of 6 by $\SI{6}{\micro\meter}$. Figure~\ref{F_Figure_01}d shows example transmission curves of full loop devices with efficiencies up to $\SI{21}{\%}$ (CaWO$_4$) and $\SI{13}{\%}$ (YVO$_4$) with full-width-half maxima in the range of 10s of nm. With this we follow the freestanding strategy from Dory et al. (2019)\cite{Dory.2019} adapted to transferable silicon photonics. Other freestanding and/or transfer strategies use photonic crystal grating coupler \cite{Ourari.2023}, Bragg grating coupler with bridges \cite{Okajima.2025, Hansen.2023, Barth.2015} or tapered fibers \cite{Dibos.2018}. In addition to the grating coupler design strategy, we employed a design layout bias method to fine-tune the transmission window of our grating couplers, thereby recovering device performance from fabrication deterioration~\cite{Sauerzopf.10312025}. 

\section{Power Dependence} \label{S_Power}
\begin{figure*}[t]
    \centering
    \includegraphics[page=1, width=\linewidth]{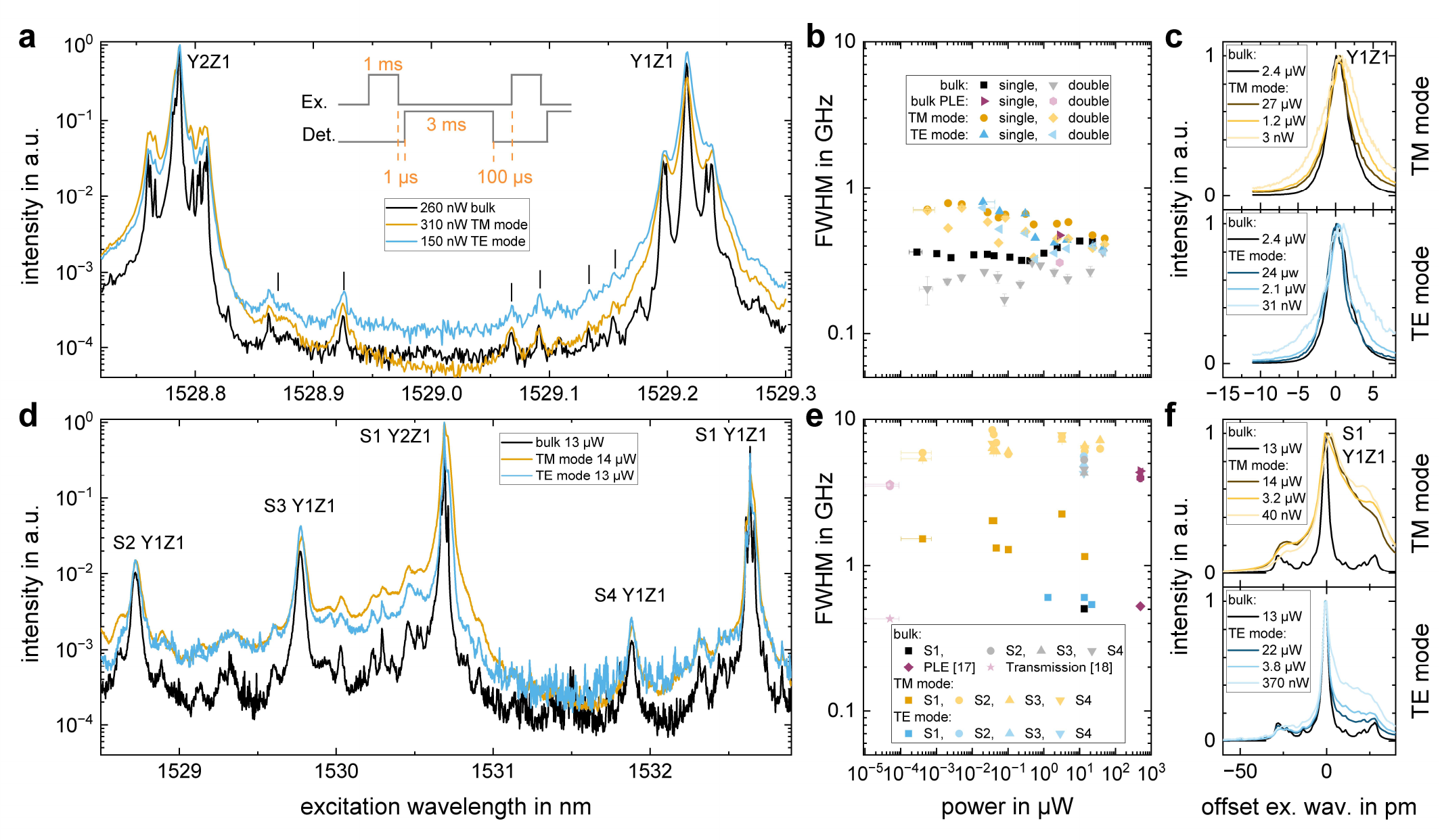}
    \caption{Comparison of pulsed excitation spectra of Er$^{3+}$:YVO$_4$ (\textbf{a, b, c}) and Er$^{3+}$:CaWO$_4$ (\textbf{d, e, f}). \textbf{a} and \textbf{d} Large excitation wavelength scans with highlighted peaks. \textbf{b} and \textbf{e} \ac{FWHM} of Lorentzian fit Y1 peaks recorded at different excitation powers. Single and double assign a single peak or double peak fit. \ac{PLE} data taken under cw excitation and recorded with a spectrometer. Transmission data recorded in \cite{Becker.10242025}. \textbf{c} and \textbf{f} High resolution pulsed excitation comparison of \ac{TM} and \ac{TE} modes at different powers with respect to the bulk peak center (ex. wav. for excitation wavelength).}
    \label{F_Figure_02}
\end{figure*}
In a first step, we investigate power and mode-dependent surface broadening of Er$^{3+}$:YVO$_4$ and Er$^{3+}$:CaWO$_4$.\\
In Figure \ref{F_Figure_02} we show this comparison of not intentionally 
terminated YVO$_4$ and CaWO$_4$. We recorded all measurements using the pulse sequence highlighted in the inset of Figure \ref{F_Figure_02}a. This includes a $\SI{1}{ms}$ excitation, a $\SI{3}{ms}$ detection (gating) pulse and optimized waiting times in-between. As a source, we used a wavelength-stabilized Toptica CTL-1500, chopped it in excitation with two AOMs and in detection with one gating AOM to avoid long recovery times of the detection SNSPDs. We cooled the samples down to $\SI{4.3}{K}$ in an Attocube attoDRY 800XS unless otherwise stated. The power noted is the measured power at a 90:10 beam splitter corrected to the power entering the cryostat. The distinction between \ac{TM} and \ac{TE} modes is summarized in Appendix \ref{S_Appendix}. \\
Figure \ref{F_Figure_02}a shows the pulsed excitation spectra of Er$^{3+}$:YVO$_4$ covering its Y2Z1 and Y1Z1 absorption lines. Overall, the spectra appear quite similar. Around the Y2Z1 and Y1Z1 absorption lines, the hyperfine levels are visible. Additionally, the background around these main peaks is elevated for \ac{TM} and \ac{TE} modes in comparison to the bulk. Finally, we highlighted dim peaks in between, which we will discuss in more detail in Section \ref{S_Temperature}. \\
Figure \ref{F_Figure_02}b shows the inhomogeneous linewidth dependency of Er$^{3+}$:YVO$_4$ with highlighted peaks in Figure \ref{F_Figure_02}c. We fitted a Lorentzian function, however, in many cases, a second Lorentzian peak was deemed necessary to better reproduce the data. Thus, we show both fits noted with single and double. While the bulk \ac{FWHM} of single fits saturates or decreases in case of double fits with decreasing power, the \ac{FWHM} of \ac{TM} and \ac{TE} mode peaks increases. The highlighted normalized peaks in \ref{F_Figure_02}c show that this behavior is not a misfit, but rather a genuine relative broadening. We attribute this contrast to the power broadening expected for bulk peaks, which is offset by the power narrowing of our surface-coupled ions to saturation. At lower powers, certain ions couple strongly to the waveguide mode, however, with increasing power, their intensity saturates. At the same time, other ions around the peak center saturate later. \\
In Figure \ref{F_Figure_02}d we present the spectra of Er$^{3+}$:CaWO$_4$. Highlighted are the sites we identified earlier \cite{Becker.2025, Becker.10242025} including the axial S1 site with its Y1Z1 and Y2Z2 peaks, as well as the Y1Z1 transitions of site S2 - S4.  In comparison, the bulk spectra show a lower background between the peaks with sharp background peaks. Additionally, around the S1 Y2Z1 transition, the \ac{TM} mode is even more elevated and more broadened than the \ac{TE} mode. \\
Figure \ref{F_Figure_02}e shows the power dependency of Lorentzian-fitted inhomogeneous broadening with highlighted examples in Figure \ref{F_Figure_02}f. Additionally, we plotted the inhomogeneous boardings from our previous studies \cite{Becker.2025, Becker.10242025}. The S2 - S4 transition show a \ac{FWHM} for bulk in the range of $\SI{4.3}{}$ - $\SI{5.3}{GHz}$, with a broadening of the \ac{TE} mode to $\SI{4.3}{}$ - $\SI{5.6}{GHz}$ and for the \ac{TM} mode to $\SI{5.3}{}$ - $\SI{8.4}{GHz}$. The S1 Y1Z1 transition, however, shows a larger dependency on the coupled mode. While the bulk \ac{FWHM} is around $\SI{0.50}{GHz}$ the \ac{TM} mode transition is broadened to $\SI{1.1}{}$ - $\SI{2.2}{GHz}$. However, the \ac{TE} mode transition \ac{FWHM} remain narrow with $\SI{0.54}{}$ - $\SI{0.60}{GHz}$.  In the close up in of both modes transitions (see Figure \ref{F_Figure_02}f), a prominent shoulder at higher wavelengths is visible in addition to its general broadening. Given these non-Lorentzian shapes, especially the \ac{FWHM} of the TM-modes, own a larger uncertainty. We discuss these fits for clarity in more detail in Appendix \ref{S_Y1_Fits}. Additionally, the measured bulk inhomogeneous broadening of S1 Y1Z1 aligns with our previously measured linewidths, whereas the Y1Z1 transition of S2-S4 appears broader.\\  
Overall, the signal of coupled modes to Er$^{3+}$ in YVO$_4$ reproduces the bulk characteristic with an elevated background and an inverse power broadening. On the other side, the modes coupling to Er$^{3+}$ in CaWO$_4$ produce also an elevated background, however, the main peaks appear with a significant shoulder and broadening up to $\SI{1.1}{}$ - $\SI{2.2}{GHz}$ for a \ac{TM} mode coupling to the S1 Y1Z1 transition.\\
To validate the role of phonon contributions to background, shoulders and broadenings, we investigate the temperature dependence in the next section.

\section{Temperature Dependence} \label{S_Temperature}
\begin{figure}[t]
    \centering
    \includegraphics[page=1, width=\linewidth, trim={0 50pt 0 0}, clip]{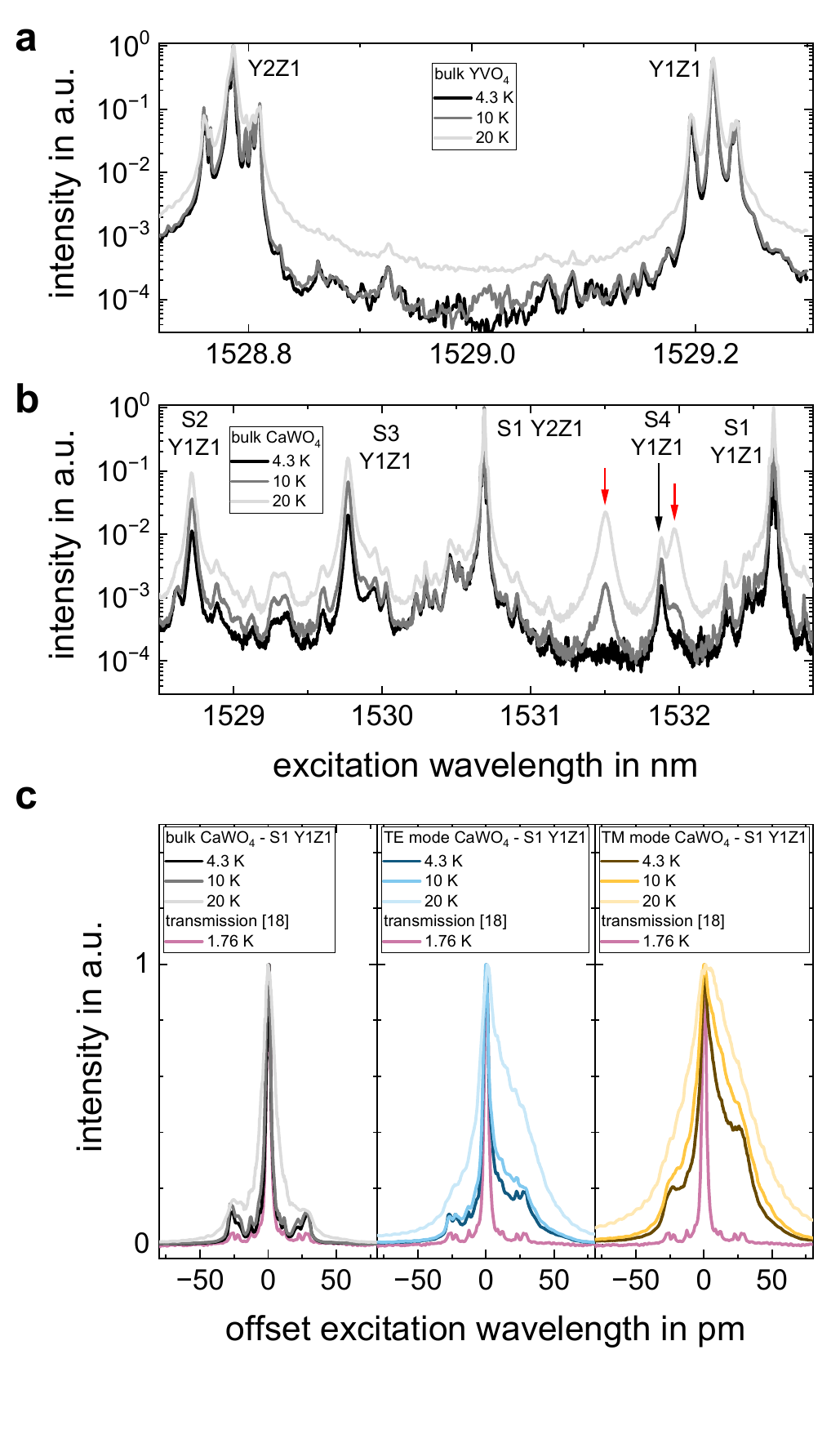}
    \caption{Crystal temperature variations probed by bulk and guided mode measurements. \textbf{a} YVO$_4$ bulk measurements at $\SI{100(2)}{\micro W}$. \textbf{b} CaWO$_4$ bulk measurements at $\SI{54.1(11)}{\micro W}$. \textbf{c} CaWO$_4$ measurements of the S1 Y1Z1 emission peak for different modes with respect to the bulk peak center.}
    \label{F_Figure_03}
\end{figure}
Figure \ref{F_Figure_03} shows the temperature influence on the spectra of Er$^{3+}$ in YVO$_4$ and CaWO$_4$. Figure \ref{F_Figure_03}a shows the influence of the bulk spectra of Er$^{3+}$:YVO$_4$. While the $\SI{4.3}{K}$ and $\SI{10}{K}$ are not significantly different, the background of the $\SI{20}{K}$ spectra is increased. This indicates that a considerable amount of phonons become activated between $\SI{10}{K}$ and $\SI{20}{K}$. What we can not see is a significant increase in the smaller background peaks. Hence, these peaks are likely not related to phonons, but rather other dilute Er$^{3+}$ defects. For instance, they could emerge from Er-Er pairs \cite{Orlovskii.2022, Shakurov.2008, GuillotNoel.2004}, other Er-impurity defects from remaining trace rare-earth ions \cite{LeDantec.2021, Buryi.2020} or other growth impurities \cite{Yokota.2025, Billaud.2025B}. \\
In Figure \ref{F_Figure_03}b we present the bulk spectra of Er$^{3+}$:CaWO$_4$. Here, the relative intensity of the site peaks increases with increasing temperature. The background at $\SI{20}{K}$ is overall elevated, however, is only increased between $\SI{1528.5}{nm}$ and $\SI{1530.0}{nm}$ at $\SI{10}{K}$. Additionally, we highlighted two new peaks that appear and increase with higher temperatures. The same two peaks appear with approximately the same energy distance on the red-shifted side of the S1 Y1Z1 peak (see Appendix \ref{S_PLE_Temperature}). Given the small energy difference of $\SI{3.1}{} - \SI{3.5}{\frac{1}{cm}}$ and $\SI{5.1}{} - \SI{5.5}{\frac{1}{cm}}$, we relate them to phonon-assisted (Anti-Stokes) absorption. However, further work would be needed to identify the reason behind the missing (blue-shifted) Stokes peaks. Furthermore, other background peaks show a comparably small temperature dependence. In Appendix \ref{S_PLE_Temperature} we relate some of them to different emission patterns observed with a spectrometer. Both together indicate that these peaks are rather additional sites than phonon-related peaks.\\
Figure \ref{F_Figure_03}c highlights the temperature dependence of bulk, \ac{TE} and \ac{TM} mode spectra and compares it to our previously measured absorption spectra \cite{Becker.10242025}. While the bulk transition merely broadens, \ac{TE} and \ac{TM} mode spectra undergo a large change. For both modes, the shoulder emerges more with an increase in temperature, especially from $\SI{10}{K}$ to $\SI{20}{K}$. As this broadening is significantly larger than what we could observe for bulk, we attribute it to the vicinity of the surface. In contrast to \cite{Billaud.2025} we can not relate it to surface strain effects due to the different thermal expansion of silicon and CaWO$_4$, as we would expect relaxing strain with temperature and no difference observable for \ac{TE} or \ac{TM} mode. In other material systems such effects are related to an unstable charge environment at the surface. For instance, for Diamond centers like NV \cite{Sangtawesin.2019, Kumar.2024}, or SiV centers \cite{Ngan.2023} at the surface, their optical performance and coherence could be improved by surface termination or passivation. This behavior is related to a more stabilized charge environment and a reduced state switching. Furthermore, for polar systems, \ac{SPhPs} are discussed in the literature. For instance, on the example of SiC pillars~\cite{Gubbin.2017} or TM polarized light coupling to SPhPs at Au gratings~\cite{Zheng.2017}.\\
The temperature and polarization dependence may be explained by electric field noise resulting from defect charge fluctuations or phonon-related processes. The Er$^{3+}$-surface defect/trap system can couple favorably, favoring an electric field normal to the surface (\ac{TM} mode), which causes the pronounced \ac{TM} shoulder. Additionally, the ionization or de-trapping that causes charge noise would be enhanced with increasing temperature. On the other side, \ac{SPhPs} are favoring coupling to normal electric fields~\cite{Zheng.2017} (\ac{TM} modes) and would also enhance with temperature according to the Bose-Einstein population. However, given the energy scale of smaller than $\SI{1}{\frac{1}{cm}}$ of the broadening/shoulder, \ac{SPhPs} are unlikely. On the other side, for a bulk \ac{REI}:CaWO$_4$ system, applied electric fields would split the two inversion symmetry-related S$_4$ sites in opposite directions. Thus, electric fields would lead to a symmetric line broadening \cite{Billaud.2025, Mims.1964, Mims.1965}.\\

To further elaborate whether this surface effect is influenced by surface charging we will study its dependence on different surface terminations in the next section. 

\section{Surface Termination Dependence} \label{S_Termination}
\begin{figure}[t]
    \centering
    \includegraphics[page=1, width=\linewidth, trim={0 50pt 0 0}, clip]{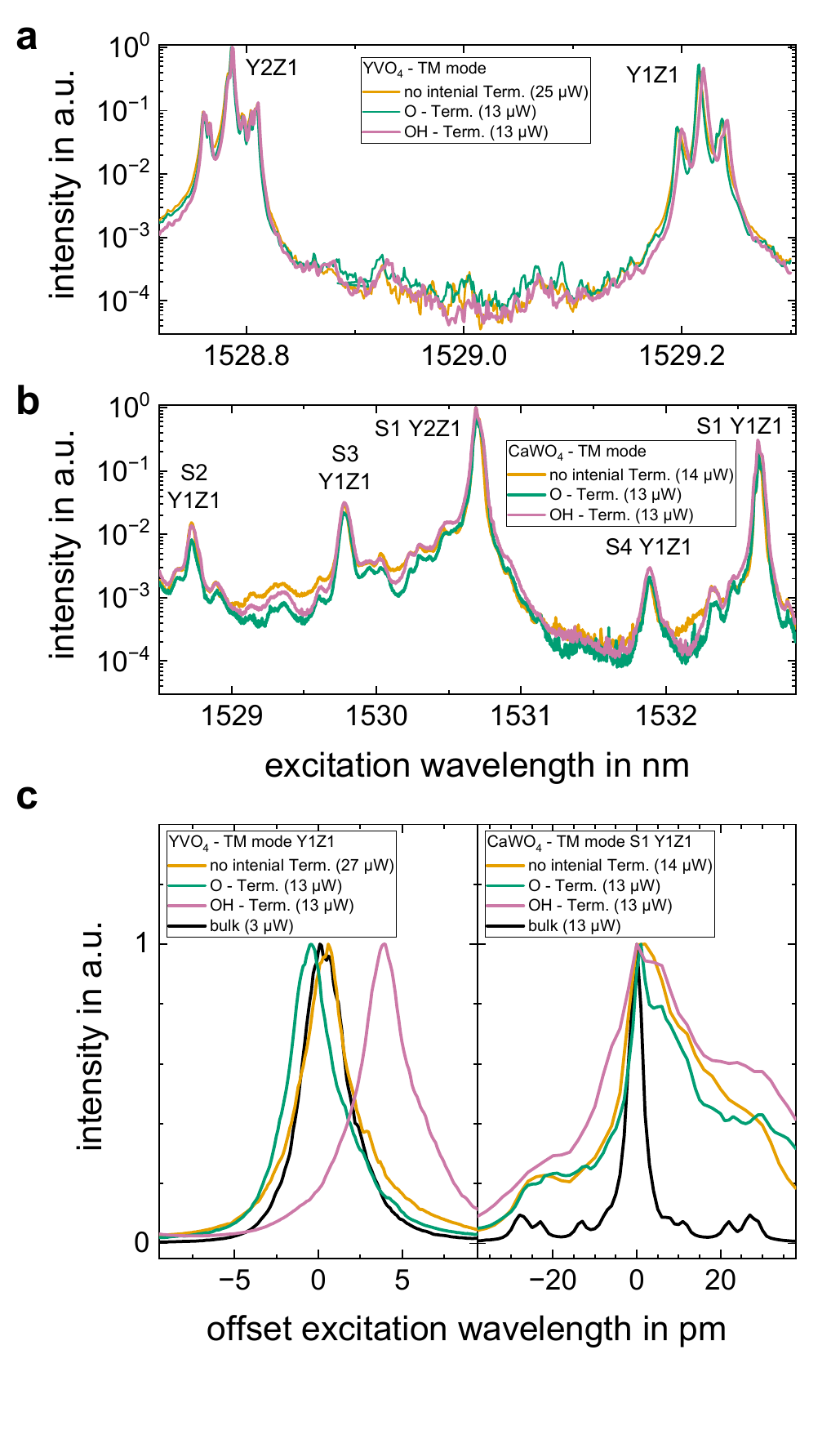}
    \caption{Comparison of TM mode coupled spectra with different surface terminations. \textbf{a} YVO$_4$ and \textbf{b} CaWO$_4$ large scans. \textbf{c} YVO$_4$ and CaWO$_4$ (S1) high resolution scans of the Y1Z1 peaks with respect to the bulk peak center.}
    \label{F_Figure_04}
\end{figure}
As we did observe the largest influence of surface effects in the \ac{TM} mode, we will restrain the termination discussion to this mode only. \\
Figure \ref{F_Figure_04} compares the \ac{TM} mode spectra of Er$^{3+}$ YVO$_4$ and CaWO$_4$ with different terminated surfaces. In detail, we compare a non-inertial terminated surface, a surface which we O-terminated, and an OH-terminated surface. The O-termination was achieved by applying a $\SI{200}{W}$ O-plasma for $\SI{2}{minutes}$. For the OH-termination, we left the crystal wetting for $\SI{30}{minutes}$ in a humid environment after the same O-plasma step. Between termination and contact of surface and \ac{PIC} approximately $\SI{10}{minutes}$ passed.\\
Figure \ref{F_Figure_04}a shows the spectra for Er$^{3+}$:YVO$_4$. The spectra show some difference around the Y1Z1 transition and at some background peaks. If we take a closer look at the Y1Z1 transition in Figure \ref{F_Figure_04}c, we observe that the O-termination produces a blue shift of approximately $\SI{1}{pm}$ and the OH-termination produces a red shift of approximately $\SI{4.5}{pm}$ in comparison to the bulk and non-inential terminated surface. Potential causes could be an electric or strain field produced by the surface termination acting at the nanoscale on the ions. In bulk Er$^{3+}$:YVO$_4$ electric and strain fields should be first-order insensitive through its high site symmetry~\cite{Xie.2021}, however, in the vicinity of the surface, local symmetries could be further restricted. Thus, the effect of electric and strain fields could be higher. Additionally, the peak shapes themselves do not seem to vary drastically.\\
Figure \ref{F_Figure_04}b presents the Er$^{3+}$:CaWO$_4$ spectra. Overall, the background of the O-terminated surface appears to be the lowest, however, besides of this, there is not much of a difference visible. Figure \ref{F_Figure_04}c shows a zoom to the S1 Y1Z1 transition. In this magnification, the phonon shoulders appear less pronounced for the O-terminated surface. For the OH-terminated surface there is also some difference compared to the non-intential terminated surface. Nevertheless, the shoulder seems to be worse than before. We conclude that an O-termination seems somewhat favorable in terms of a reduced shoulder. This agrees with color centers near the surface in diamond, which optically perform better under a more stabilized/defined charge environment~\cite{Sangtawesin.2019, Kumar.2024, Ngan.2023}.

\section{Conclusion}
In contrast to our initial assumption that the deeper addressed Er$^{3+}$ ions by \ac{TM} modes would exhibit better optical performances, we found that Er$^{3+}$:CaWO$_4$ ions show a large shoulder/broadening coupled by \ac{TM} modes. At the same time, \ac{TE} modes exhibit optical transitions that closely represent bulk absorption. The comparison platform Er$^{3+}$:YVO$_4$ does not show this surface effect. Additionally, the broadening/shoulder shows a temperature and surface termination dependency. Given the sensitivity of color centers in diamond to surface charges~\cite{Sangtawesin.2019, Kumar.2024, Ngan.2023} and the not charge neutral replacement of Ca$^{2+}$ with Er$^{3+}$ in CaWO$_4$ it seems likely, that the surface Er$^{3+}$ is prone to charge fluctuations. Additionally, the termination sensitivity of Er$^{3+}$:YVO$_4$ supports a picture of surface charge dependence. However, charge fluctuations should only cause symmetric broadening through the pseudo-Stark splitting of the two inversion symmetry-related axial S1 sites in CaWO$_4$~\cite{Billaud.2025, Mims.1964}.\\
Certainly, further work is needed to elucidate why electric field noise from surface charges might cause a preferential energy reduction of the S1 Y1Z1 transition in Er$^{3+}$:CaWO$_4$, rather than symmetrical broadening and why TE fields seem to offer the less affected coupling strategy.  

Another next step for quantum applications could be comparing the coherence times of bulk and surface ions, as well as those of different surface terminations. 
 
\section{Acknowledgements}\label{S_Acknowledge}
We gratefully acknowledge support
from the German Federal Ministry of Research, Technology and Space (BMFTR) via the funding program Photonics Research Germany (project MOQUA, contract number 13N14846) and project 6G-life, the Bavarian State Ministry for Science and Arts (StMWK) via projects EQAP and NEQUS, the Bavarian Ministry of Economic Affairs (StMWi) via project 6GQT, as well as from the German Research Foundation (DFG) under Germany’s Excellence Strategy EXC-2111 (390814868) and projects PQET (INST 95/1654-1) and MQCL (INST 95/1720-1).

\section{Data Availability}
The data that support the findings of this study are available from the corresponding author upon reasonable request.
\section{Conflicts of Interest}
The authors declare no conflicts of interest.

\section{Appendix}\label{S_Appendix}
\subsection{Polarization Dependent Transmission}\label{S_PolPlots}
\begin{figure}[t]
    \centering
    \includegraphics[page=1, width=\linewidth, trim={0 100pt 0 0}, clip]{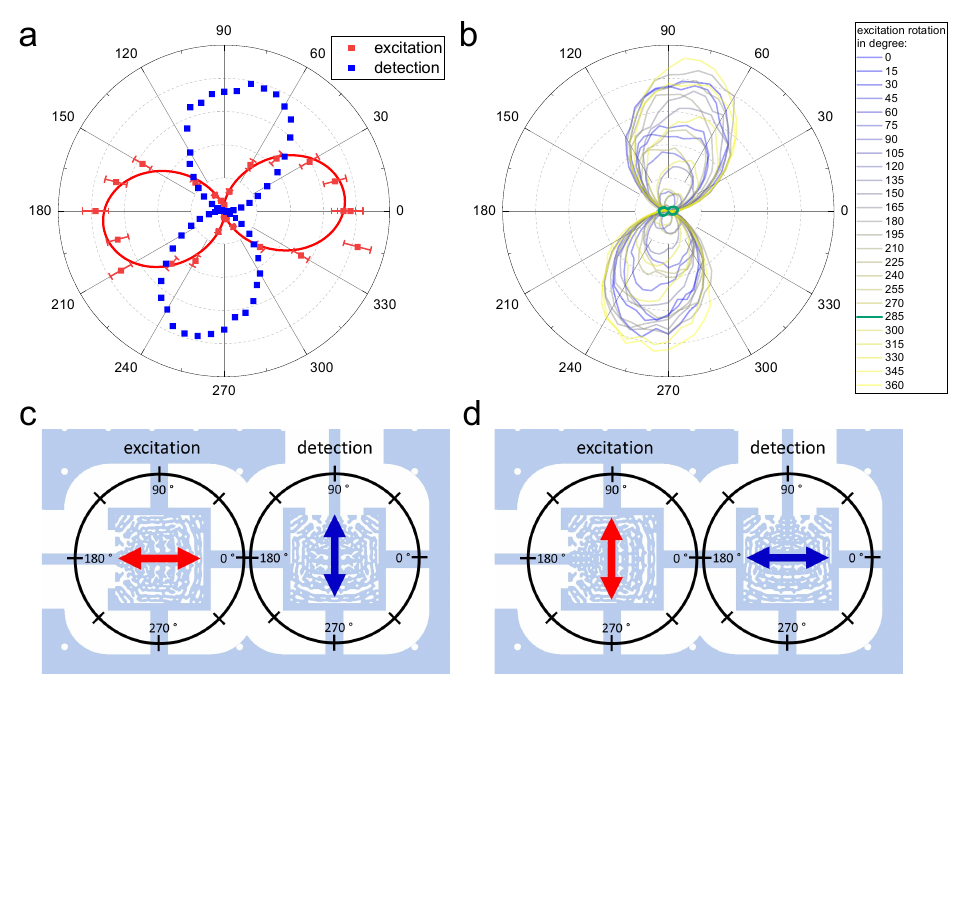}
    \caption{Polarization dependence of excitation and detection. \textbf{a} Polarization orientations. \textbf{b} Polarization Excitation sweep (raw data for subfigure a) \textbf{c} Supported main transmission polarization (TM mode) with indicated polarizations \textbf{d} Supported minor transmission polarization (TE mode) with indicated polarizations.}
    \label{F_Figure_A_01}
\end{figure}
In this section, we present an example of the polarization dependence of our grating couplers. Figure \ref{F_Figure_A_01}a shows the resulting main polarization of detection and excitation. For these measurements, the laser transmission was recorded by a spectrometer as the polarizing optics of excitation and detection were rotated. The laser was kept at a wavelength of $\SI{1530}{nm}$ for these measurements. For the detection polarization, only the detection polarizing optics were rotated. For the excitation polarization, we choose a set of orientations for which we recorded a full detection polarization scan. We show the raw data of this excitation polarization scan with rotated detection polarizations for every increment in Figure \ref{F_Figure_A_01}b. The amplitude of every excitation increment is represented as one data point in Figure \ref{F_Figure_A_01}a. In sum, excitation and detection polarization plots are almost orthogonal. We attribute the misaligned axis to a misaligned linear polarization axis in the free space setup. Figure \ref{F_Figure_A_01}c highlights this main polarization orientation of detection and excitation, which aligns with the polarization in our simulation. In Figure \ref{F_Figure_A_01}b, the detection polarizations are mainly oriented along the $\SI{90}{\degree}$ - $\SI{270}{\degree}$ axis for the majority of excitation orientations. However, around an excitation orientation of $\SI{90}{\degree}$ or $\SI{270}{\degree}$ (highlighted $\SI{285}{\degree}$), the detection polarization flips to the opposite polarization. Figure \ref{F_Figure_A_01}d highlights this extreme. For grating couplers supporting TE modes, we would choose this source polarization, however, it seems that our TM mode optimized grating couplers allow some coupling to TE modes as well. From this data, we get an approximate ratio of TM:TE mode amplitude of 25:1. We can exclude leakage from the excitation for the TE modes, as in this orientation, excitation and detection are orthogonal to each other.
\subsection{Relaxation Time}\label{S_T1}
\begin{figure}[t]
    \centering
    \includegraphics[page=1, width=\linewidth]{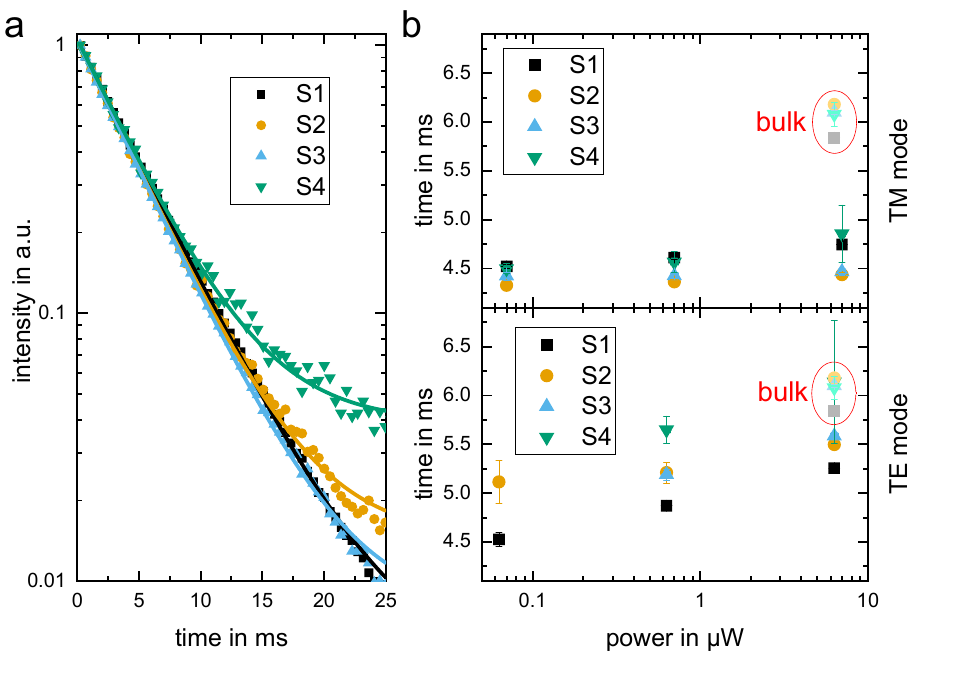}
    \caption{Relaxation times of coupled modes. \textbf{a} Example relaxation measurement of TM mode coupled Er$^{3+}$ in CaWO$_4$. \textbf{b} Power dependent relaxation times of bulk, TM and TE mode.}
    \label{F_Figure_A_02}
\end{figure}
Another key indication of whether we couple with a waveguide mode to the Er$^{3+}$ is the lifetime. The density of states that we couple to should differ from bulk measurements and between different supported modes. Thus, also the small Purcell enhancement from waveguides should lead to mode-dependent lifetimes.\\
Figure \ref{F_Figure_A_02} shows depopulation times observed when coupled to the TM mode, TE mode, and bulk for our different sites in CaWO$_4$. Figure \ref{F_Figure_A_02}a highlights a typical decay together with a fit. In Figure \ref{F_Figure_A_02}b we present the resulting depopulation times. Both modes appear to have different relaxation times compared to bulk measurements. Additionally, the TE mode exhibits timings closer to those of the bulk than the TM mode. However, all three measurements are different.\\ \\

Finally, both polarization dependence and Purcell-enhanced relaxation times allow us to distinguish the predominantly supported TM mode and a minorly supported TE mode. 
\subsection{Y1Z1 Peaks Fit}\label{S_Y1_Fits}
\begin{figure}[t]
    \centering
    \includegraphics[page=1, width=\linewidth]{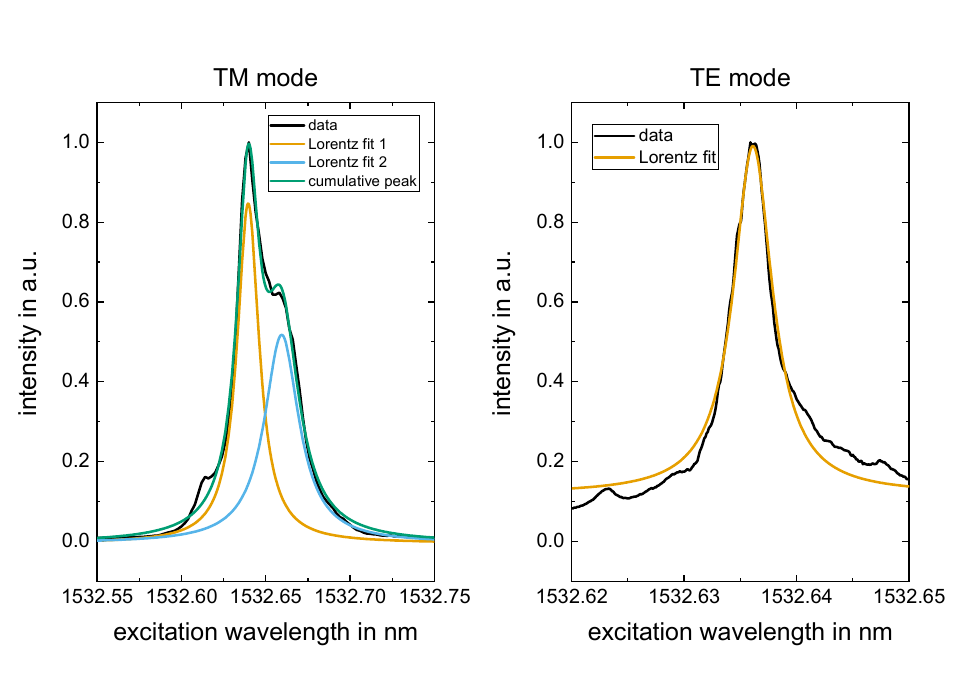}
    \caption{Example Lorentzian fits on S1 Y1Z1 peaks of Er$^{3+}$:CaWO$_4$ for TM and TE mode.}
    \label{F_Figure_A_03}
\end{figure}
In this section, we discuss the trustworthiness of our Lorentzian fits on Er$^{3+}$:CaWO$_4$ for TM and TE mode coupling. Figure \ref{F_Figure_A_03} shows a typical fit for the S1 Y1Z1 peak. While the TE mode can be reproduced by a single Lorentzian fit quite well, the TM mode peak needs two fitted peaks for some accuracy. Thus, the TM inhomogeneous linewidth provides only an orientation and would be more than twice as large if calculated from the raw data instead of the fits.
\subsection{Phonons} \label{S_PLE_Temperature}
\begin{figure}[t]
    \centering
    \includegraphics[page=1, width=\linewidth]{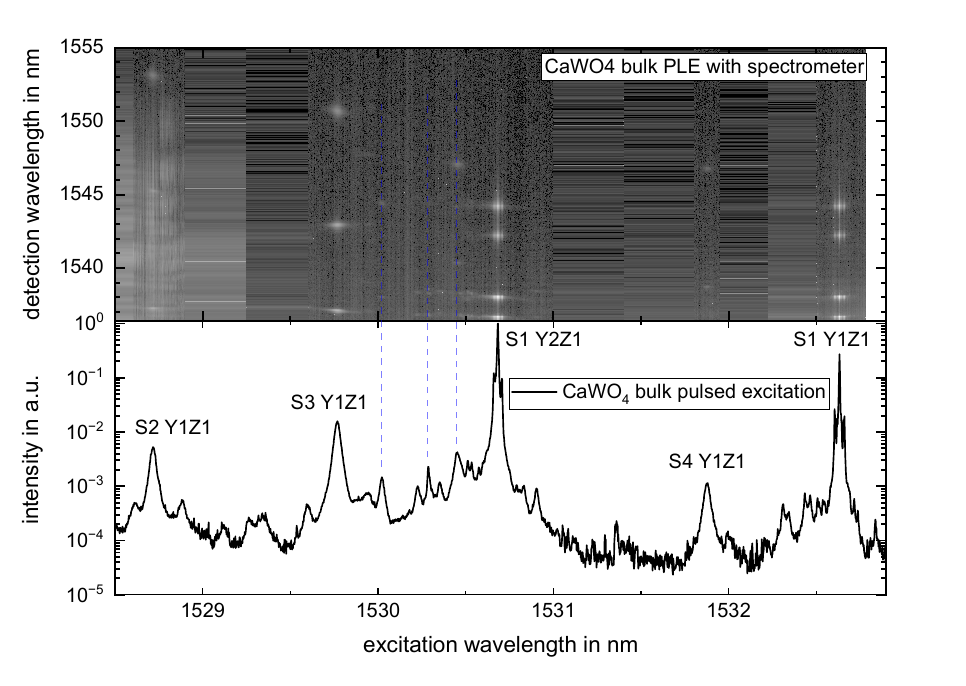}
    \caption{Comparison of \ac{PLE} measured with a spectrometer in detection and the pulsed excitation on bulk CaWO$_4$ both at $\SI{4.3}{K}$. The AOM shift of $\SI{450}{MHz}$ is corrected for the pulsed excitation spectra.}
    \label{F_Figure_A_04}
\end{figure}
Figure \ref{F_Figure_04} shows a comparison of our earlier measured \ac{PLE} dataset from \cite{Becker.2025} with the current bulk pulsed-excitation scans. While all main sites (S1 - S4) can be clearly attributed to the spectrometer measurements, the background peaks between S3 Y1Z1 and S1 Y2Z1 also show emissions at different wavelengths. Hence, in combination with their low temperature dependence (see Section \ref{S_Temperature}), we believe they are other sites rather than phonon-related peaks. 
\begin{figure}[t]
    \centering
    \includegraphics[page=1, width=\linewidth]{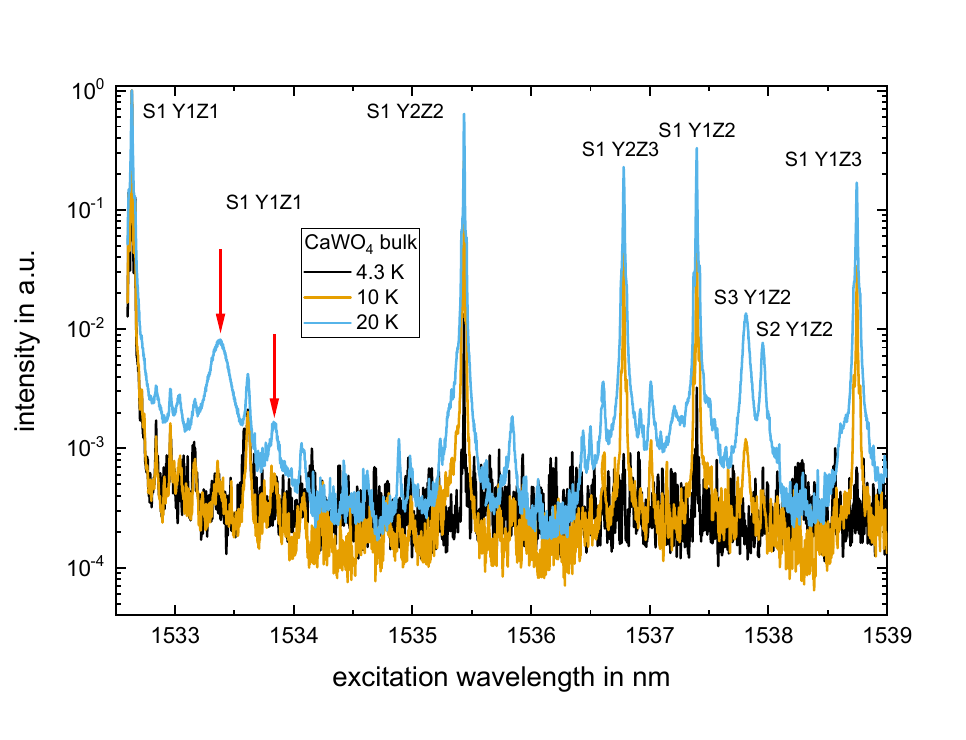}
    \caption{Pulsed excitation scan to higher wavelengths on bulk CaWO$_4$ at different sample temperatures.}
    \label{F_Figure_A_05}
\end{figure}
Figure \ref{F_Figure_A_05} shows a scan on bulk CaWO$_4$ at different temperatures. We labelled visible stark transitions and highlighted the discussed sharp anti-Stokes resonances. While at $\SI{4.3}{K}$ mainly the S1 Y1Z1 transition is visible, with increasing temperatures, also other Stark levels and phonon-related transitions appear. The increase of the Z2 and Z3 Stark field level population following Boltzmann's law causes an appearance and an intensity increase of these transitions with temperature. The other peaks appearing with temperature are also likely phonon-related peaks.

\providecommand{\noopsort}[1]{}\providecommand{\singleletter}[1]{#1}%

\end{document}